\documentclass[3p,sort&compress]{elsarticle}
\usepackage{amsmath,amssymb}
\usepackage{graphicx} 
\usepackage{url} 
\usepackage{subfigure} 

\usepackage{color}

\newcommand{\Teq}{T_{\mathrm{eq}}}
\newcommand{\aeq}{a_{\mathrm{eq}}}
\newcommand{\UR}{U_{\mathrm{R}}}
\newcommand{\uR}{u_{\mathrm{R}}} 
\newcommand{\teq}{t_{\mathrm{eq}}}
\newcommand{\vecasc}{\vec{a}^{\mathrm{sc}}_N}
\newcommand{\vecone}{\vec{1}_N}
\newcommand{\Ksc}{K^{\mathrm{sc}}}
\newcommand{\ER}{E_{\mathrm{R}}}

\newcommand{\ST}{S_{q}}
\newcommand{\SR}{S_{\mathrm{R}q}}

\newcommand{\zetaB}[3]{\zeta_{\mathrm{B}}(#1,#2|#3)}

\newcommand{\hH}{\hat{H}}
\newcommand{\hrho}{\hat{\rho}} 
\newcommand{\Tr}{\mathrm{Tr}}
\newcommand{\hdelta}[1]{\hat{\delta}_{#1}}

\begin{document}

\begin{frontmatter}
  \title{Description using equilibrium temperature in the canonical ensemble
    within the framework of the Tsallis statistics employing the conventional expectation value}
\journal{}
\author{Masamichi Ishihara\corref{cor}}
\cortext[cor]{Corresponding author.} 
\ead{email: masamichi.ishihara.research@gmail.com}
\address{Department of Economics, Faculty of Economics, Chiba Keizai University, Chiba, 263-0021, JAPAN}


\begin{abstract}
  We studied the thermodynamic quantities and the probability distribution,
  expressing the probability distribution as a function of the energy,
  in the canonical ensemble within the framework of the Tsallis statistics, 
  which is characterized by the entropic parameter $q$, employing the conventional expectation value (the linear average).
  We treated the power-law-like distribution. 
  The equilibrium temperature, which is often called the physical temperature, was employed, 
  and the probability distribution described with the equilibrium temperature was derived. 
  The Tsallis statistics represented by the equilibrium temperature was applied to $N$ harmonic oscillators,
  where $N$ is the number of the oscillators. 
  The expressions of the energy, the Tsallis entropy, and the heat capacity were obtained. 
  The expressions of these quantities and the expression of the probability distribution were obtained
  when the differences between adjacent energy levels are the same.
  These quantities and the distributions were numerically calculated.
  The $q$ dependences of the energy, the R\'enyi entropy, and the heat capacity are weak.
  In contrast, the Tsallis entropy depends on $q$.
  The probability distribution as a function of the energy depends on $N$ and $q$.
  The results provide a basis for describing power-law-like phenomena in the Tsallis statistics.
  The present formulation is expected to apply to various phenomena,
  because the harmonic oscillator plays a fundamental role in describing classical and quantum systems.
\end{abstract}


\end{frontmatter}

\section{Introduction}


The extension of the Boltzmann--Gibbs statistics has been developed.
The Tsallis statistics is one parameter extension of the Boltzmann--Gibbs statistics \cite{Tsallis:Book:2010}.
The Tsallis statistics is based on the definition of the Tsallis entropy and that of the expectation value \cite{Tsallis:PhysicaA:1998}. 
The Tsallis statistics employing the conventional expectation value (the linear average) is often called the Tsallis-1 statistics. 
The Tsallis statistics employing the unnormalized $q$-expectation value is often called the Tsallis-2 statistics. 
The Tsallis statistics employing the normalized $q$-expectation value (the escort average) is often called the Tsallis-3 statistics.


The maximum entropy principle \cite{Jaynes:1957} is applied to obtain the probability.
The obtained probability contains the Tsallis entropy and the energy in the Tsallis-1 statistics.
It was found that
the expectation value of the Tsallis-1 statistics and the expectation value of the Tsallis-3 statistics are invariant to energy shifts
\cite{Parvan:PLA:2006, Parvan:PhysicaA:2022, Ishihara:2024:EPJP139}.
This invariance is preferable to describe the system,
because this invariance implies that the physical quantities do not depend on the zero-point of the energy.


Two temperatures appear in the Tsallis statistics.
One is the temperature $T$ as the inverse of the Lagrange multiplier.
The other is the equilibrium temperature $\Teq$
\cite{Imdiker:EPJC:2023, Ishihara:EPJP:2023:1, Ishihara:EPJP:2023:2, Ishihara:2024:EPJP139}
which is often called the physical temperature
\cite{Abe-PLA:2001, Kalyana:2000, S.Abe:physicaA:2001, Aragao:2003, Ruthotto:2003, Toral:2003, Suyari:2006, Ishihara:phi4, Ishihara:free-field, Ishihara:2024:EPJP139, Ignatyuk_2024}.
The equilibrium temperature is defined with the physical quantities of each subsystem, 
and the equilibrium temperature is the same across the subsystems.
Therefore, the description using the equilibrium temperature may be preferable. 


Heat capacity has an important role in the Tsallis statistics.
The temperature fluctuation is related to the heat capacity and leads to the $q$-exponential type distribution \cite{Wilk:EPJA40:2009}.
The Tsallis distribution also appears in the system of the constant heat capacity \cite{Wada2003},  
and the condition between the entropic parameter and the heat capacity was shown \cite{Ishihara:EPJP:2023:2}.
It is worthwhile to study the heat capacity in the Tsallis statistics.


It is important to study the system composed of harmonic oscillators in the Tsallis statistics,  
because harmonic oscillators are basic to describe physical systems.
Thermodynamic quantities of the system are expressed by the Barns zeta function \cite{Ruijsenaars:2000, Kirsten:2010}
which is expanded with the Hurwitz zeta function \cite{Elizalde1989, Oprisan, Ishihara:2024:EPJP139}.
The number of constituents is limited from the requirement for the Barnes zeta function \cite{Ruijsenaars:2000}. 
Such limitations have been revealed in the previous studies \cite{Abe-PLA:2001, Lenzi:PLA:2001, Ishihara:EPJB:2022, Ishihara:EPJB:2023, Ishihara:2024:EPJP139}.
A transverse momentum spectrum in high-energy collisions was proposed by using a harmonic oscillator \cite{Bhattacharyya:Tsallis-1:2025}.
The study of the system composed of harmonic oscillators is meaningful in various branches of physics.

The formalism of the Tsallis statistics employing the conventional expectation value and the equilibrium temperature $\Teq$ 
should be useful to describe power-law-like phenomena.
However, the Tsallis statistics employing the conventional expectation value and the temperature $T$
was applied to the system of harmonic oscillators in Ref.~\cite{Ishihara:2024:EPJP139}.
The conventional expectation value is a reasonable candidate of the expectation value,
because the expectation value is generally used to obtain the expectation values of physical quantities.
The introduction of $\Teq$ provides a natural representation of the probability.
It was found that the standard thermometer measures $\Teq$ in the Tsallis statistics \cite{Kalyana:2000}.
It was also clarified in the microcanonical ensemble that  
$\Teq$ corresponds to the temperature appearing in the Boltzmann--Gibbs statistics
and that the free energy described using $\Teq$ corresponds to the free energy in the Boltzmann--Gibbs statistics \cite{Toral:2003}.
These results indicate a possible equivalence between the thermodynamic quantity expressed using the equilibrium temperature
and the quantity expressed using the temperature appearing in the Boltzmann--Gibbs statistics in the canonical ensemble.
The $\Teq$-dependence of a quantity should be studied when the temperature is measured by a standard thermometer.
Therefore, it is worthwhile to study the system in the Tsallis-1 statistics employing the equilibrium temperature.

In this paper, we derive the probability in the Tsallis-1 statistics, adopting the equilibrium temperature.
We obtain the expressions of the energy, the Tsallis entropy, and the heat capacity in the system of harmonic oscillators.
We also calculate the probability distribution, expressing the probability distribution as a function of the energy.
We show that the entropy is eliminated by introducing the equilibrium temperature $\Teq$,
whereas previous studies have shown that the entropy is explicitly contained in the probability described using the temperature $T$
in the Tsallis-1 statistics \cite{Parvan:PLA:2006,Parvan:PhysicaA:2022,Ishihara:2024:EPJP139}.  
The quantities as functions of $\Teq$ are calculated in this study.
In contrast, the quantities as functions of $T$ were calculated in the previous studies.
We find that the energy, the R\'enyi entropy, and the heat capacity barely depend on $N$ and $q$ and that the Tsallis entropy depends on $N$ and $q$,
where $N$ is the number of the oscillators and $q$ is the entropic parameter. 
We show that the probability distribution depends explicitly on $N$ and $q$ even when the equilibrium temperature is adopted. 
These results offer a framework for describing power-law-like phenomena.

This paper is organized as follows.
In Section~\ref{sec:Tsallis-1:prob:eq-tempe}, we derive the probability represented by the equilibrium temperature
in the Tsallis statistics employing the conventional expectation value.
In Section~\ref{sec:HO}, we employ multiple harmonic oscillators.
We calculate the thermodynamic quantities and
the probability distribution, expressing the probability distribution as a function of the energy. 
In Section~\ref{sec:numerical_calc}, we calculate numerically the quantities and the distribution. 
The last section is assigned for discussion and conclusions. 
The derivation of the probability $p_i$ and the derivation of the density operator $\hrho$
in the Tsallis statistics employing the conventional expectation value are given in \ref{appendix:derivations}.

\section{Probability represented by the equilibrium temperature in the Tsallis statistics employing the conventional expectation value}
\label{sec:Tsallis-1:prob:eq-tempe}
\subsection{The Tsallis statistics employing the conventional expectation value}
The Tsallis statistics characterized by the entropic parameter $q$ is based on the definition of the Tsallis entropy and that of the expectation value.
The Tsallis entropy $\ST$ is given by
\begin{align}
  \ST = \frac{\left( \displaystyle\sum_i (p_i)^q \right) - 1 }{1-q},  
  \label{def:S:Tsallis}
\end{align}
where
$p_i$ is the probability of being in state $i$.
We employ the conventional expectation value. The expectation value of a quantity $A$ is given by
\begin{align}
\langle A \rangle = \sum_i A_i p_i,  
\end{align}
where $A_i$ is the value of $A$ in state $i$.

The probability $p_i$ in classical statistical physics is obtained by applying the maximum entropy principle in the canonical ensemble. 
The functional $I_{\mathrm{cl}}$ with both the normalization condition and the energy constraint is defined by
\begin{align}
  I_{\mathrm{cl}} = \ST - \alpha \left[ \left( \sum_i p_i\right)  - 1 \right] - \beta \left[ \left(\sum_i p_i E_i \right) - U \right],
  \label{eqn:functional}
\end{align}
where $E_i$ is the value of the energy in state $i$ and $U$ is the energy.
The parameters, $\alpha$ and $\beta$, are the Lagrange multipliers.  
The probability in the Tsallis statistics employing the conventional expectation value 
(the Tsallis-1 statistics) \cite{Parvan:PLA:2006, Parvan:PhysicaA:2022, Ishihara:2024:EPJP139} is
\begin{align}
p_i = \left[ 1 + (1-q) \ST + \left( \frac{1-q}{q} \right) \beta (E_i - U) \right]^{1/(q-1)} . 
\end{align}
The probability follows a power-law-like distribution when $q$ is less than one.
It may be worthwhile to mention that the functional given by Eq.~\eqref{eqn:functional} is different from that given in the reference \cite{Tsallis:1988}.
The density operator $\hrho$ plays an essential role in quantum statistical physics. 
The operator $\hrho$ in the present statistics can be derived in the same manner:
\begin{align}
  \hrho = \left[1 + (1-q) \ST + \left(\frac{1-q}{q}\right) \beta (\hH-U) \right]^{1/(q-1)} ,
\end{align}
where $\hH$ is the Hamiltonian and $\ST$ is the Tsallis entropy.
The Tsallis entropy in quantum statistical physics is defined by $\ST = (\Tr(\hrho^q) - 1)/(1-q)$.
The derivation of the probability $p_i$ is given in \ref{Tsallis-1-Parvan:probability}
and the derivation of the density operator $\hrho$ is given in \ref{Tsallis-1-Parvan:density-operator}.

The R\'enyi entropy $\SR$ often appears in various branches of science.
The R\'enyi entropy is defined by 
\begin{align}
\SR = \frac{\ln \left( \displaystyle\sum_i (p_i)^q \right)}{1-q}.
\label{def:S:Renyi}
\end{align}
The R\'enyi entropy is related to the Tsallis entropy.
From Eqs.~\eqref{def:S:Tsallis} and \eqref{def:S:Renyi}, we have
\begin{align}
\SR = \frac{\ln\left(1+(1-q) \ST \right)}{1-q}.
\end{align}
Therefore, the R\'enyi entropy is obtained when the Tsallis entropy is obtained.

\subsection{Probability represented by the equilibrium temperature}

The equilibrium temperature, which is often called the physical temperature, was introduced in the Tsallis statistics.
The quantity $\ST^{\mathrm{A}}$ is the Tsallis entropy of subsystem A, $\ST^{\mathrm{B}}$ is that of subsystem $B$,
and $\ST^{\mathrm{A+B}}$ is that of the system composed of A and B.
When the pseudo-additivity of the Tsallis entropy $\ST^{\mathrm{A+B}} = \ST^{\mathrm{A}} + \ST^{\mathrm{B}} + (1-q) \ST^{\mathrm{A}} \ST^{\mathrm{B}}$
and the additivity of the energy $U^{A+B} = U^{A} + U^{B}$ hold, 
the requirement for the entropy $\delta \ST^{\mathrm{A+B}} =0$ and that for energy $\delta U^{\mathrm{A+B}} = 0$ give
\begin{align}
  \frac{1}{1+(1-q) \ST^{\mathrm{A}}} \frac{\partial \ST^{\mathrm{A}}}{\partial U^{\mathrm{A}}}
  = \frac{1}{1+(1-q) \ST^{\mathrm{B}}} \frac{\partial \ST^{\mathrm{B}}}{\partial U^{\mathrm{B}}} .
\end{align}
Therefore, the equilibrium temperature $\Teq$ is introduced by 
\begin{align}
  \frac{1}{\Teq} = \frac{1}{1+(1-q) \ST} \frac{\partial \ST}{\partial U}.
\end{align}
The Lagrange multiplier $\beta$ is related to the derivative of $\ST$ with respect to $U$: $\beta = {\partial \ST}/{\partial U}$.
Therefore, we have
\begin{align}
  \frac{1}{\Teq} = \frac{\beta}{R},
\end{align}
where $R$ is defined by $R = 1 +(1-q) \ST$.

The probability $p_i$ is represented by $R$ and $\Teq$ for $R > 0$:
\begin{align}
p_i = R^{ \frac{1}{q-1} } \left[ 1 + \left(\frac{1-q}{q} \right) \frac{E_i - U}{\Teq} \right]^{\frac{1}{q-1}} . 
\end{align}
It is possible to remove $R$ from the expression of $p_i$ by using the relation $R = \sum_i (p_i)^q$:
\begin{align*}
  R^{\frac{1}{q-1}}
    = \frac{1}{\displaystyle\sum_j \left[1 + \left(\frac{1-q}{q}\right) \left( \frac{E_j - U}{\Teq} \right) \right]^{\frac{q}{q-1}} }.
\end{align*}  
We obtain the expression of $p_i$:
\begin{align}
  p_i = \frac{\left[ 1 + \left(\frac{1-q}{q} \right) \left(\frac{E_i - U}{\Teq}\right) \right]^{\frac{1}{q-1}} }
  {\displaystyle\sum_j \left[1 + \left(\frac{1-q}{q}\right) \left(\frac{E_j - U}{\Teq}\right) \right]^{\frac{q}{q-1}} }.
  \label{p_i}
\end{align}
The probability $p_i$ contains the energy $U$.
Therefore, the expectation value of the energy gives the self-consistent equation.
The entropy $S_q$ is not contained in $p_i$ when the equilibrium temperature is adopted.

\section{Description of the system composed of harmonic oscillators in the Tsallis statistics employing the conventional expectation value}
\label{sec:HO}

\subsection{Energy of the system and the Barnes zeta function}
The system of harmonic oscillators is important to describe physical systems.
The total energy $E$ of the system is given by
\begin{align*}
E = \sum_j (a_j n_j + b_j),  
\end{align*}
where $a_j$ and $b_j$ are the parameters of the oscillator numbered $j$.
The quantity $n_j$ is a non-negative integer. 
We apply the Tsallis-1 statistics to describe the system, adopting the equilibrium temperature.

The Barnes zeta function $\zeta_B$ appears in the calculations.
The function $\zeta_B$ is defined by
\begin{align}
\zeta_B(s,a | a_1, \cdots, a_N) = \sum_{n_1=0,\cdots,n_N=0}^{\infty} \frac{1}{(a+a_1n_1+a_2n_2+\cdots+a_N n_N)^{s}} ,
\end{align}
where $a$ and $a_j$ are positive.
The parameter $s$ is required to satisfy the inequality $s>N$.

We introduce the expression $\vec{a}_N = (a_1, a_2, \cdots, a_N)$ for simplicity. 
We employ the following expression with the representation $\vec{a}_N$:
\begin{align}
  \zeta_B(s,a | \vec{a}_N) = \zeta_B(s,a | a_1, \cdots, a_N) . 
\end{align}
Similarly, we introduce the expression $\vec{1}_N = (1, 1, \cdots, 1)$.
The function $\zeta_B(s,a | \vec{1}_N)$ indicates
\begin{align}
\zeta_B(s,a | \vec{1}_N) = \sum_{n_1=0,\cdots,n_N=0}^{\infty} \frac{1}{(a+n_1+n_2+\cdots+n_N)^{s}} .
\end{align}

The Barnes zeta function can be represented by the Hurwitz zeta function \cite{Oprisan, Elizalde1989, Ishihara:2024:EPJP139}.
We use the relation between the Barnes zeta function and the Hurwitz zeta function in the numerical calculations.

\subsection{The expressions of the physical quantities}
The energy, the Tsallis entropy, and the heat capacity are calculated in the Tsallis-1 statistics for $q<1$.
These quantities are calculated with the probability $p_i$ given by Eq.~\eqref{p_i}.
It is easily found that the quantities except for the energy $U$ do not contain the zero-point energy $\sum_j b_j$.  
We introduce $\UR$ by
\begin{align}
  \UR = U -\sum_j b_j . 
\end{align}
The heat capacity $C$ is defined by
\begin{align}
  C = \frac{\partial U}{\partial \Teq}. 
\end{align}
The above heat capacity is defined using the equilibrium temperature. 
We introduce the following scaled quantities:
\begin{subequations}
\begin{align}
  &\uR = \UR / a_1 ,\\
  &\teq = \Teq/a_1 ,\\
  &\vecasc = (a_1/a_1, a_2/a_1, \cdots, a_N/a_1) .
\end{align}
\end{subequations}

The scaled energy $\uR$, the Tsallis entropy $\ST$, and the heat capacity $C$ are given by
\begin{subequations}
\begin{align}
  &\uR =
  \left(\frac{q\teq}{1-q}\right) \left\{
  1 + \left[ \left(\frac{1-q}{q}\right)\uR - \teq\right]
  \frac{\zetaB{\frac{1}{1-q}}{\teq-\left(\frac{1-q}{q}\right)\uR}{\left(\frac{1-q}{q}\right)\vecasc}}
       {\zetaB{\frac{q}{1-q}}{\teq-\left(\frac{1-q}{q}\right)\uR}{\left(\frac{1-q}{q}\right)\vecasc}}
  \right\} , 
  \label{Osc:scaled_ene}\\
  &\ST = \frac{(\teq)^q \left[\zetaB{\frac{q}{1-q}}{\teq-\left(\frac{1-q}{q}\right)\uR}{\left(\frac{1-q}{q}\right)\vecasc}\right]^{1-q} - 1}{1-q} ,
  \label{Osc:entropy} \\
  &C = \frac{\uR - \left(\frac{q}{1-q}\right) (\teq)^2 \Ksc}{\teq ( 1 - \teq \Ksc)} ,
  \label{Osc:heat_capacity} \\
  &\Ksc = 
  \frac{\zetaB{\frac{1}{1-q}}{\teq-\left(\frac{1-q}{q}\right)\uR}{\left(\frac{1-q}{q}\right)\vecasc}}
       {\zetaB{\frac{q}{1-q}}{\teq-\left(\frac{1-q}{q}\right)\uR}{\left(\frac{1-q}{q}\right)\vecasc}}
       + \frac{1}{1-q} \left[ \left(\frac{1-q}{q}\right) \uR - \teq \right]
       \nonumber \\
       &\quad\qquad \times
       \left\{
       \frac{\zetaB{\frac{2-q}{1-q}}{\teq-\left(\frac{1-q}{q}\right)\uR}{\left(\frac{1-q}{q}\right)\vecasc}}
            {\zetaB{\frac{q}{1-q}}{\teq-\left(\frac{1-q}{q}\right)\uR}{\left(\frac{1-q}{q}\right)\vecasc}}
       - q\left[\frac{\zetaB{\frac{1}{1-q}}{\teq-\left(\frac{1-q}{q}\right)\uR}{\left(\frac{1-q}{q}\right)\vecasc}}
         {\zetaB{\frac{q}{1-q}}{\teq-\left(\frac{1-q}{q}\right)\uR}{\left(\frac{1-q}{q}\right)\vecasc}} \right]^2
       \right\} ,
  \label{Osc:Ksc} 
\end{align}
\end{subequations}
where $\kappa \vec{a}_N$ in the Barnes zeta function represents $(\kappa a_1, \kappa a_2, \cdots, \kappa a_N)$.
The scaled energy can be obtained by solving Eq.~\eqref{Osc:scaled_ene}. 
To solve Eq.~\eqref{Osc:scaled_ene}, it is better to introduce the new variable $x$ defined by
\begin{align}
  x = \frac{q \teq}{1-q} - \uR.
  \label{eqn:x}
\end{align}
The Tsallis entropy and the heat capacity can be calculated using the solution of Eq.~\eqref{Osc:scaled_ene}.
The variable $x$ should be positive from the requirement for the Barnes zeta function. 
This requirement gives the limitation:
\begin{align}
  \uR < \left(\frac{q}{1-q}\right) \teq.  
\end{align}

\subsection{The thermodynamic quantities and the probability distribution in the system with uniform differences between adjacent energy levels}

Hereafter, we treat the case that $a_1, a_2, \cdots, a_N$ are all equal to $\aeq$.
From Eqs.~\eqref{Osc:scaled_ene}, \eqref{Osc:entropy}, \eqref{Osc:heat_capacity}, and \eqref{Osc:Ksc},
the expressions of the scaled energy, the Tsallis entropy, and the heat capacity are given by
\begin{subequations}
\begin{align}
  &\uR =
  \left(\frac{q\teq}{1-q}\right) \left\{
  1 + \left[ \uR - \frac{q \teq}{1-q}\right] 
  \frac{\zetaB{\frac{1}{1-q}}{\frac{q \teq}{1-q}-\uR}{\vecone}}
       {\zetaB{\frac{q}{1-q}}{\frac{q \teq}{1-q}-\uR}{\vecone}}
  \right\} , 
  \label{Osc:scaled_ene:equal}\\
  &\ST = \frac{\left(\frac{q\teq}{1-q}\right)^q \left[\zetaB{\frac{q}{1-q}}{\frac{q \teq}{1-q}-\uR}{\vecone}\right]^{1-q} - 1}{1-q} ,
  \label{Osc:Tsallis_Entropy:equal}\\
  &C = \frac{\uR - \left(\frac{q}{1-q}\right) (\teq)^2 \Ksc}{\teq ( 1 - \teq \Ksc)} ,
  \label{Osc:Heat_Capacity}\\
  &\Ksc =
  \left( \frac{q}{1-q} \right)
  \frac{\zetaB{\frac{1}{1-q}}{\frac{q \teq}{1-q}-\uR}{\vecone}}
       {\zetaB{\frac{q}{1-q}}{\frac{q \teq}{1-q}-\uR}{\vecone}}
       + \left( \frac{q}{(1-q)^2} \right) \left(\uR - \frac{q\teq}{1-q} \right) 
       \nonumber \\ & \qquad\qquad \times
       \left\{
       \frac{\zetaB{\frac{2-q}{1-q}}{\frac{q \teq}{1-q}-\uR}{\vecone}}
            {\zetaB{\frac{q}{1-q}}{\frac{q \teq}{1-q}-\uR}{\vecone}}
       - q\left[\frac{\zetaB{\frac{1}{1-q}}{\frac{q \teq}{1-q}-\uR}{\vecone}}
         {\zetaB{\frac{q}{1-q}}{\frac{q \teq}{1-q}-\uR}{\vecone}} \right]^2
       \right\} . 
\end{align}
\end{subequations}

Equations~\eqref{Osc:scaled_ene:equal}, \eqref{Osc:Tsallis_Entropy:equal}, and \eqref{Osc:Heat_Capacity} are simplified using Eq.~\eqref{eqn:x}:
\begin{subequations}
  \begin{align}
    &\left(\frac{q \teq}{1-q}\right)
    \frac{\zetaB{\frac{1}{1-q}}{x}{\vecone}}{\zetaB{\frac{q}{1-q}}{x}{\vecone}} - 1 = 0, \label{eqn:x:U} \\
    &\ST = \frac{\left(\frac{q \teq}{1-q}\right)^q \left[\zetaB{\frac{q}{1-q}}{x}{\vecone}\right]^{1-q}-1}{1-q}, \label{eqn:x:Sq}\\
    &C = \frac{q \teq}{1-q} - \frac{x}{\teq (1-\teq \Ksc)}. \label{eqn:x:C}
  \end{align}
\end{subequations}
Equation~\eqref{eqn:x:U} is the equation of $x$, where $x$ should be positive. 
We obtain the scaled energy $\uR$ using the solution $x$. 
We have the Tsallis entropy $\ST$ and the heat capacity $C$,
substituting the solution $x$ into Eqs.~\eqref{eqn:x:Sq} and \eqref{eqn:x:C}.
We also calculate the R\'enyi entropy $\SR$ from the Tsallis entropy $\ST$.

It is possible to obtain the probability distribution as a function of the energy. 
The energy $E$ of $N$ harmonic oscillators for $n_1 + n_2 + \cdots + n_N = M$ is given by 
\begin{align}
E \equiv E_N (M) = \aeq M + \sum_{j=1}^N b_j,   
\end{align}
where $M$ is a non-negative integer:  
The number of the states with the energy $E$ is given by
\begin{align}
g_N(E) = \left( \begin{array}{c} M+N-1 \\ N-1 \end{array} \right) .  
\end{align}
We define the energy $\ER$ by 
\begin{align}
  \ER = E - \sum_{j=1}^N b_j.  
\end{align}    
Therefore, we have 
\begin{align}
  E - U = \ER - \UR = \aeq M - \UR.
\end{align}    
We have the probability distribution $f_N(E)$ using the probability given by Eq.~\eqref{p_i}.
We use the integer $M$ instead of $E$.
We have 
\begin{align}
  f_N(E) = \frac{(M+N-1)!}{M! (N-1)!} \frac{[(q\teq)/(1-q) - \uR + M]^{1/(q-1)}}{\zetaB{\frac{q}{(1-q)}}{\frac{q\teq}{(1-q)}-\uR}{\vecone}} .
\label{eqn:probdist_of_E}
\end{align}
As shown in Eq.~\eqref{eqn:probdist_of_E},
the probability distribution $f_N(E)$ does not contain zero-point energy $\sum_{j=1}^{N} b_j$. 
This distribution is obtained using the solution of Eq.~\eqref{eqn:x:U}.

The Boltzmann--Gibbs limit $q \rightarrow 1$ of $f_N(E)$ is
\begin{align}
  \lim_{q \rightarrow 1} f_N(E)
  = \frac{(M+N-1)!}{M! (N-1)!} \left(\frac{\exp(1/\teq)}{\exp(1/\teq) -1}\right)^{-N} \exp(-M/\teq) .
\label{eqn:prob_ene_BGlim}
\end{align}
The equation $\sum_{M=0}^{\infty} \lim_{q \rightarrow 1} f_N(E) = 1$ is easily shown by using the following relation:
\begin{align}
  (1-x)^{-(N+1)}
  = \sum_{M=0}^{\infty} \left(\begin{array}{c} M+N \\ M \end{array} \right) x^M
  = \sum_{M=0}^{\infty} \left(\begin{array}{c} M+N \\ N \end{array} \right) x^M, 
  \qquad |x| < 1 . 
\end{align}

\section{Numerical calculations in the system with uniform differences between adjacent energy levels} 
\label{sec:numerical_calc}
In this section, we treat the case that $a_1, a_2, \cdots, a_N$ are all equal to $\aeq$.
We attempt to calculate thermodynamic quantities and the probability distributions numerically 
in the Tsallis-1 statistics for various values of the parameters. 
The value of the parameter $q$ is set to satisfy the inequality $N/(N+1)<q<1$.

\subsection{Thermodynamic quantities}
We calculate the energy, the Tsallis entropy, and the heat capacity in the Tsallis-1 statistics.
Equation~\eqref{eqn:x:U} is numerically solved for $x$.
We obtain the scaled energy $\uR$, the Tsallis entropy $\ST$, the R\'enyi entropy $\SR$, and the heat capacity $C$.

\begin{figure}
  \centering
  \subfigure[The scaled energies]
            {
              \includegraphics[width=0.45\textwidth]{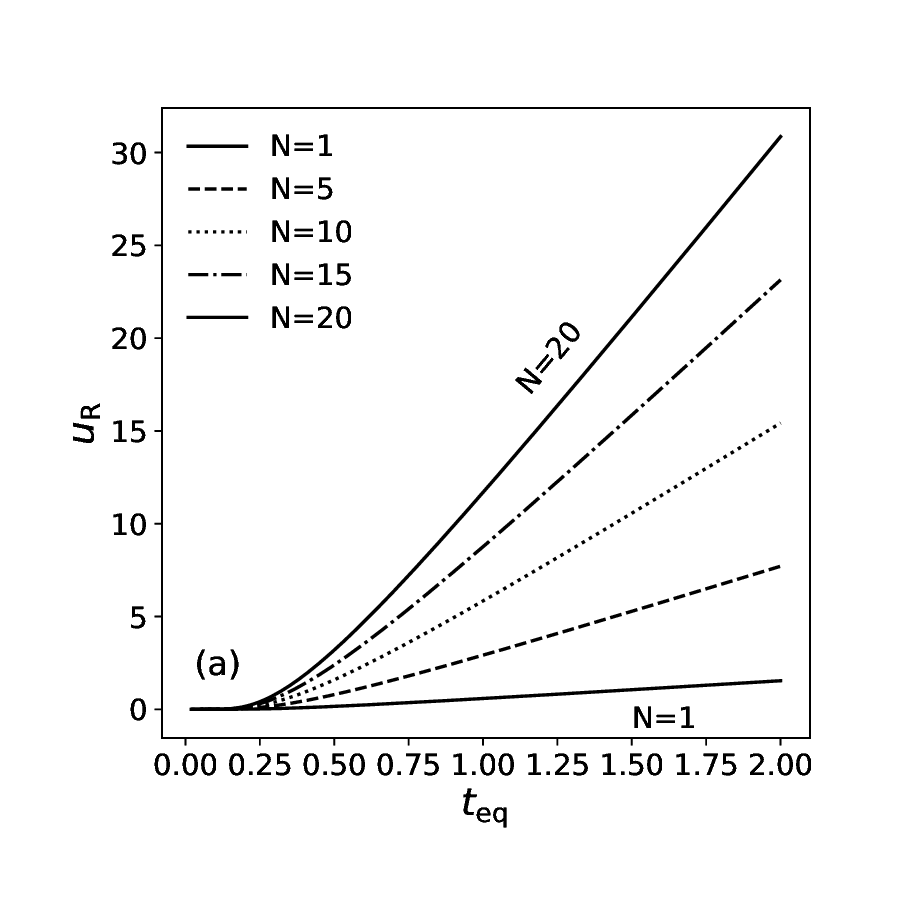}
              \label{fig:ene:Ndep}
            }
            \hfill
  \subfigure[The scaled energies divided by $N$]
  {
    \includegraphics[width=0.45\textwidth]{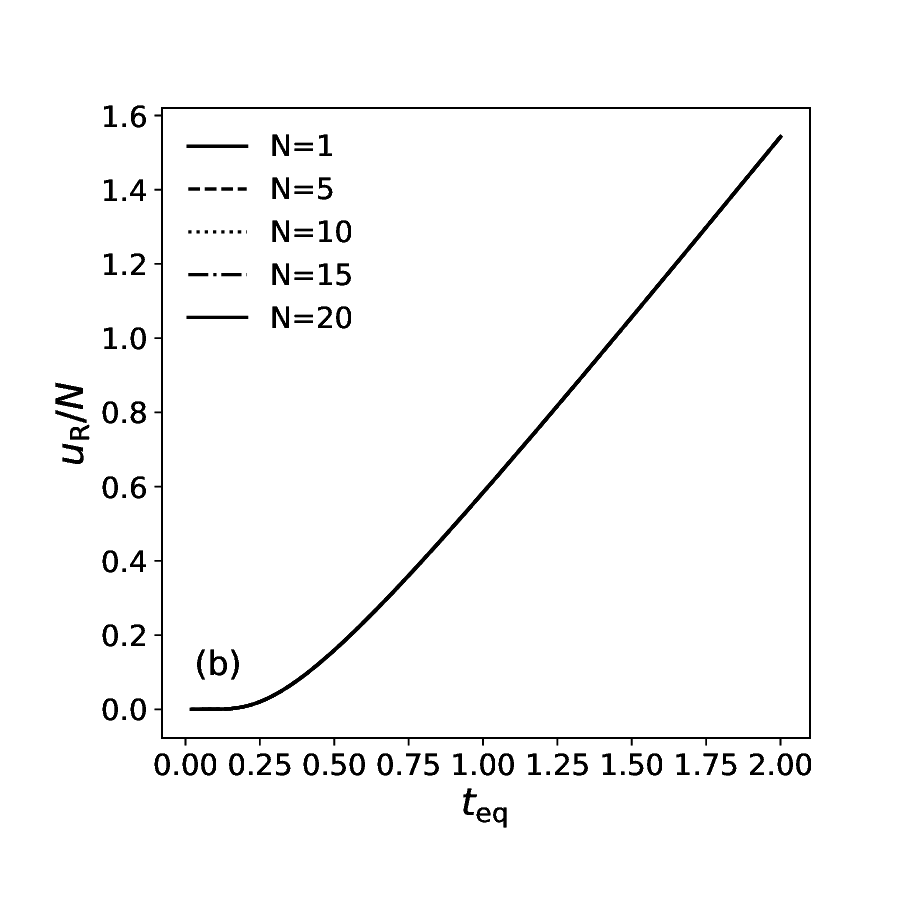}
    \label{fig:ene:Ndep:rescaled}
  }
  \subfigure[The scaled energies divided by $N$ in the narrow range]{
    \includegraphics[width=0.45\textwidth]{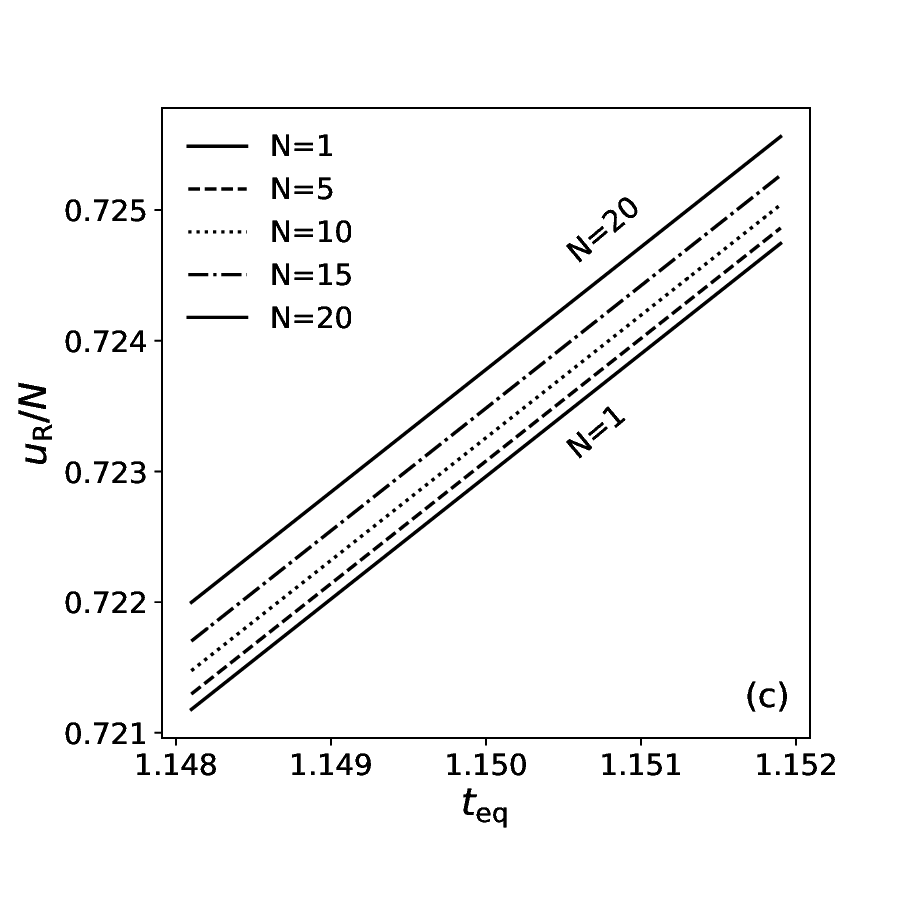}
    \label{fig:ene:Ndep:rescaled:prec}
  }
  \caption{The scaled energies $\uR$ and the scaled energies divided by $N$ ($\uR/N$), as functions of $\teq$ at $q=0.98$ for $N=1, 5, 10, 15$, and $20$.}
\end{figure}

\begin{figure}
  \centering
  \includegraphics[width=0.45\textwidth]{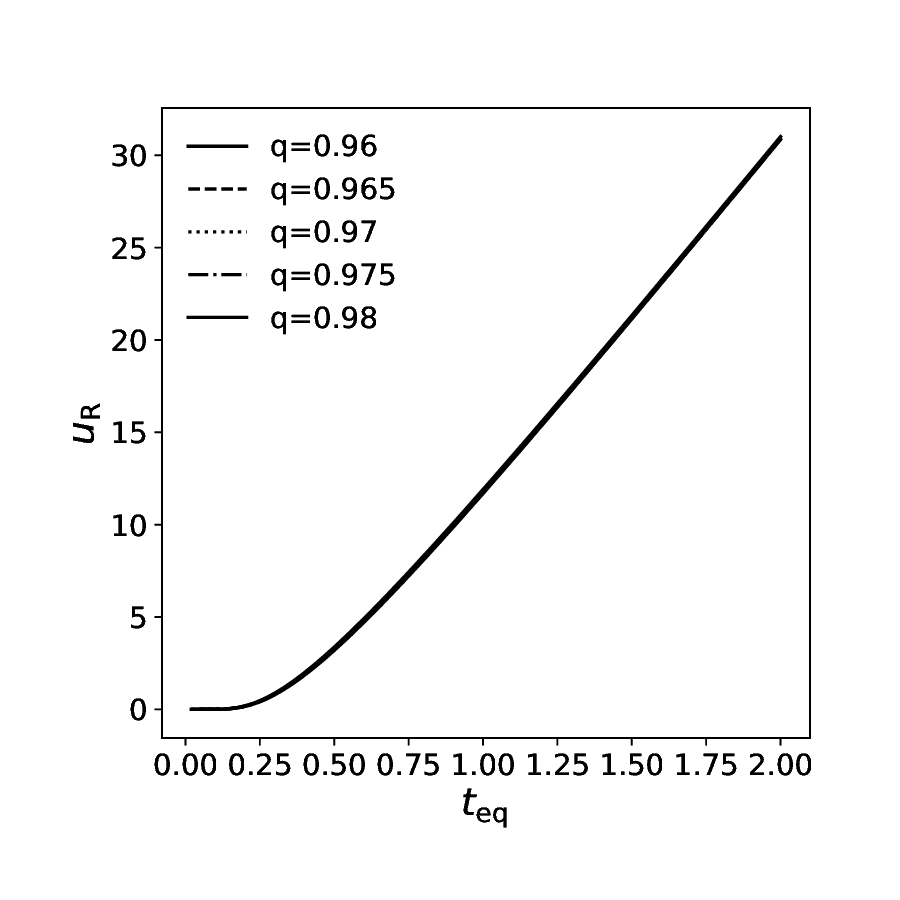}
  \caption{The scaled energies $\uR$ as functions of $\teq$ at $N=20$ for $q=0.96, 0.965, 0.97, 0.975$, and $0.98$.}
  \label{fig:ene:qdep}
\end{figure}

First, we calculate the scaled energy $\uR$ numerically.  
Figure~\ref{fig:ene:Ndep} shows the scaled energies $\uR$ at $q=0.98$ as functions of $\teq$ for $N=1, 5, 10, 15$, and $20$.
The scaled energy $\uR$ increases with $N$. 
Figure~\ref{fig:ene:Ndep:rescaled}~\ref{fig:ene:Ndep:rescaled:prec}
shows the scaled energies per oscillator $\uR/N$ at $q=0.98$ as functions of $\teq$ for $N=1, 5, 10, 15$, and $20$.
The range of Figure~\ref{fig:ene:Ndep:rescaled:prec} is narrow. 
It is found from these figures that the $N$ dependence of $\uR/N$ is exceedingly weak.
This indicates that the energy $\UR$ for $N$ oscillators is approximately $N$ times the energy for a single oscillator. 
Figure~\ref{fig:ene:qdep} shows the scaled energies $\uR$ at $N=20$ as functions of $\teq$ for $q=0.96, 0.965, 0.97, 0.975$, and $0.98$.
The $q$ dependence of $\uR$ is also exceedingly weak.

\begin{figure}
  \centering
  \subfigure[The Tsallis entropies]
  {
    \includegraphics[width=0.45\textwidth]{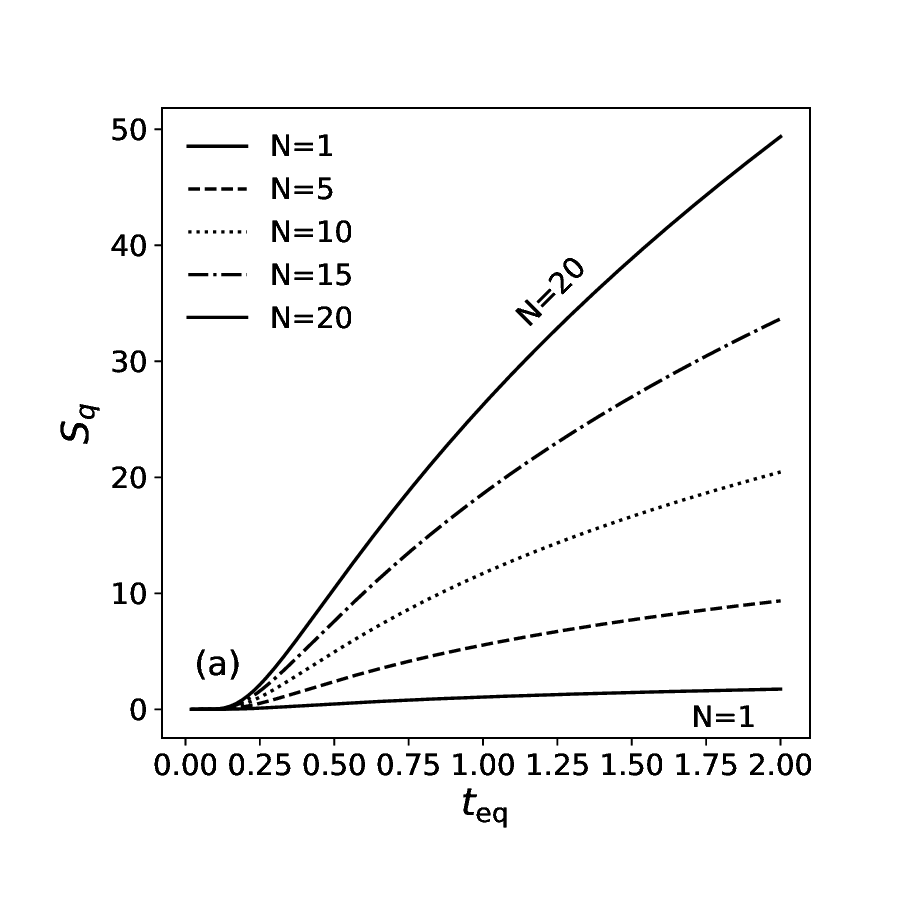}
    \label{fig:TsallisEntropy:Ndep}
  }
  \hfill
  \subfigure[The Tsallis entropies divided by $N$]
  {
    \includegraphics[width=0.45\textwidth]{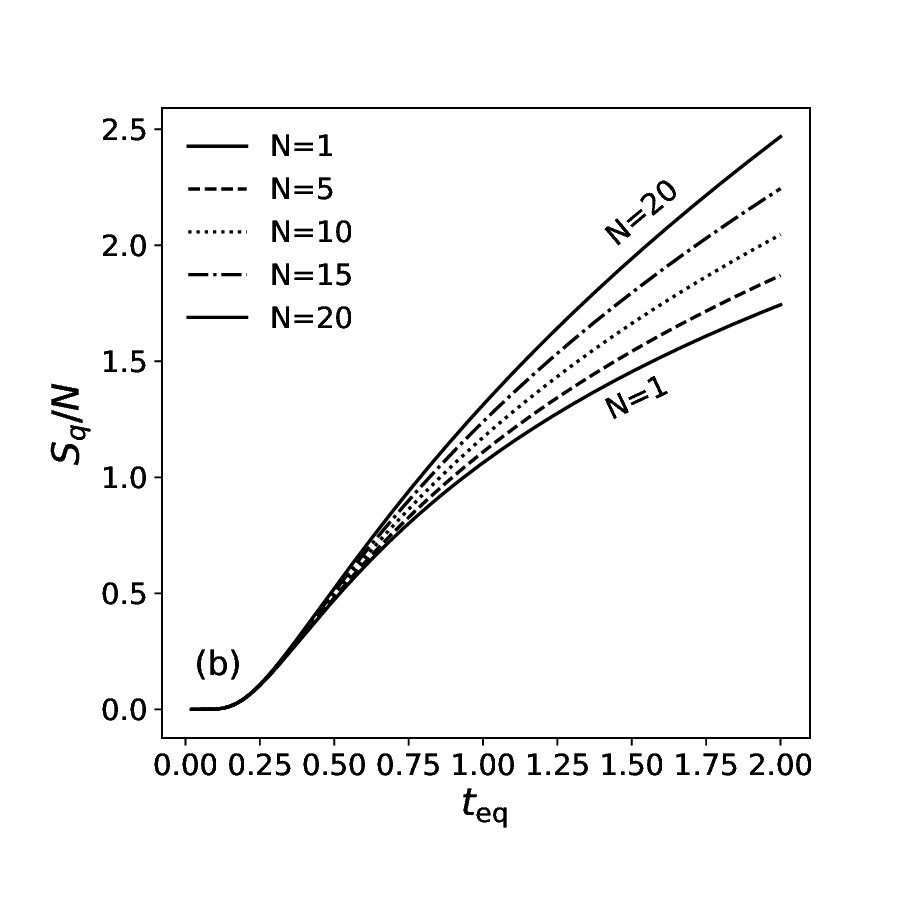}
    \label{fig:TsallisEntropy:Ndep:rescaled}
  }
  \caption{The Tsallis entropies $\ST$ and the Tsallis entropies divided by $N$ ($\ST/N$), as functions of $\teq$ at $q=0.98$ for $N=1, 5, 10, 15$, and $20$.}
\end{figure}

\begin{figure}
  \centering
  \includegraphics[width=0.45\textwidth]{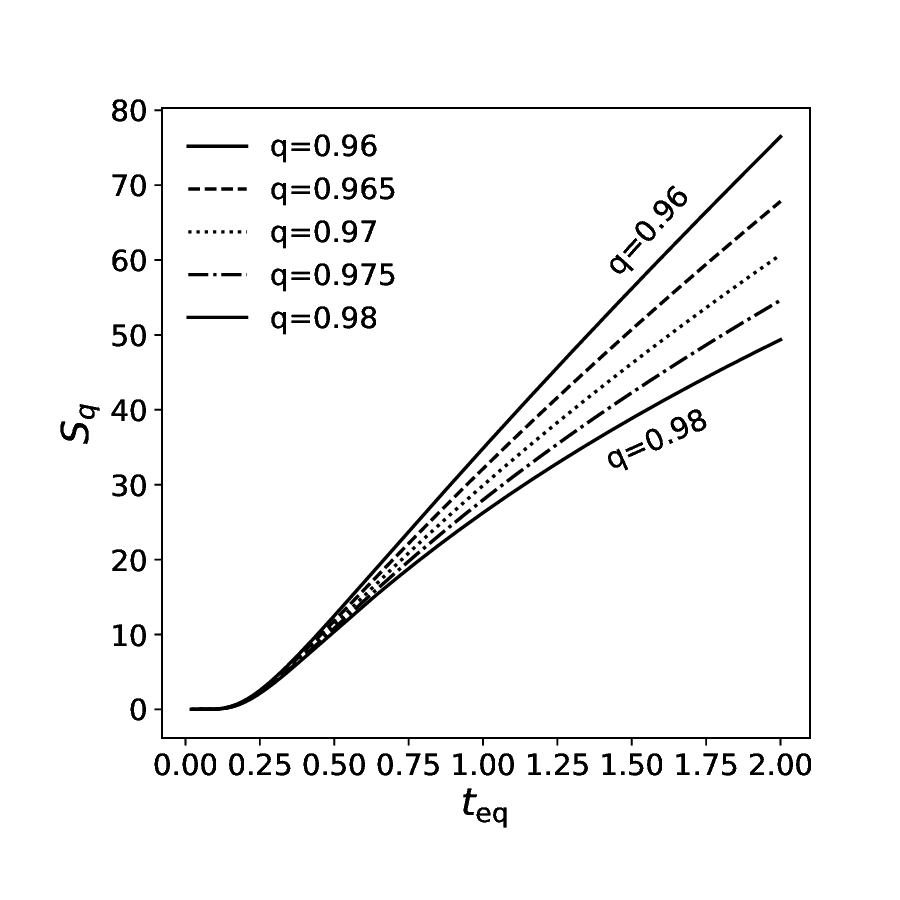}
  \caption{The Tsallis entropies $\ST$ as functions of $\teq$ at $N=20$ for $q=0.96, 0.965, 0.97, 0.975$, and $0.98$.}
  \label{fig:TsallisEntropy:qdep}
\end{figure}

Second, we calculate the Tsallis entropy $\ST$ numerically.  
Figure~\ref{fig:TsallisEntropy:Ndep} shows the Tsallis entropies $\ST$ at $q=0.98$ as functions of $\teq$ for $N=1, 5, 10, 15$, and $20$.
As expected, the Tsallis entropy increases with $N$. 
Figure~\ref{fig:TsallisEntropy:Ndep:rescaled} shows the Tsallis entropies per oscillator $\ST/N$ at $q=0.98$ as functions of $\teq$ for $N=1, 5, 10, 15$, and $20$.
The quantity $\ST/N$ increases with $N$.
Therefore, the value of $\ST/N$ for a positive integer $N$ ($N \ge 2$) is not equal to $N$ times the value of $\ST/N$ at $N=1$. 
Figure~\ref{fig:TsallisEntropy:qdep} shows the Tsallis entropies $\ST$ at $N=20$ as functions of $\teq$ for $q=0.96, 0.965, 0.97, 0.975$, and $0.98$.
The Tsallis entropy decreases with $q$. 

\begin{figure}
  \centering
  \subfigure[The R\'enyi entropies]
  {
    \includegraphics[width=0.45\textwidth]{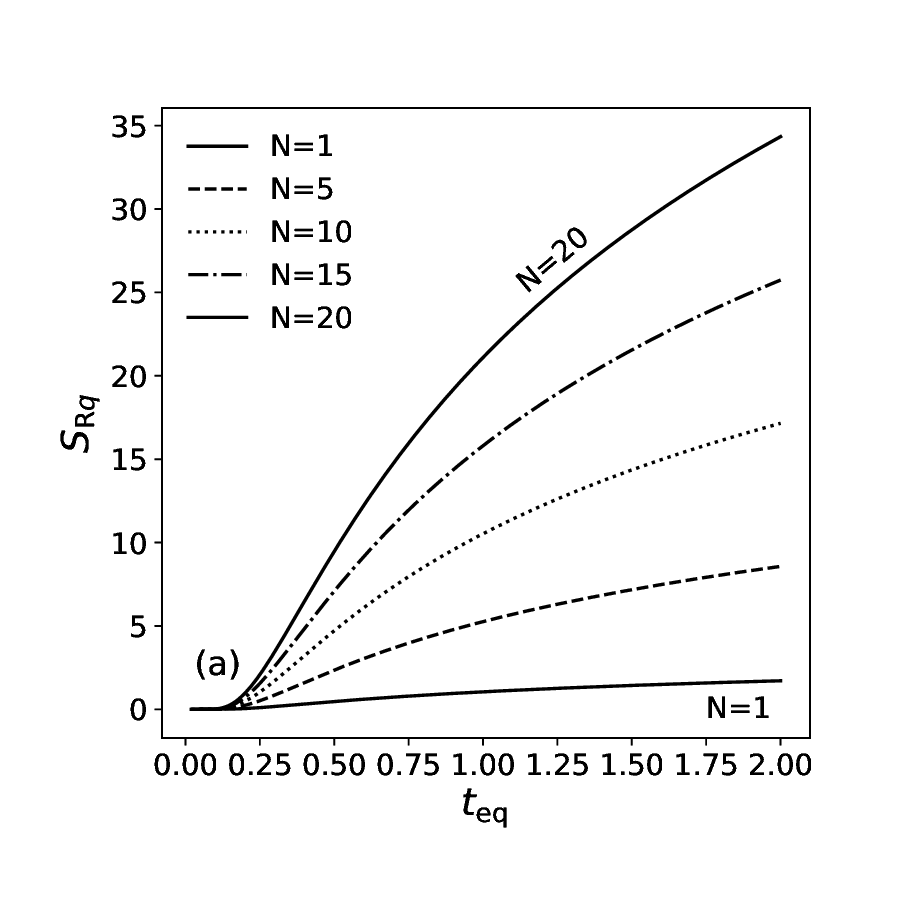}
    \label{fig:RenyiEntropy:Ndep}
  }
  \hfill
  \subfigure[The R\'enyi entropies divided by $N$]
  {
    \includegraphics[width=0.45\textwidth]{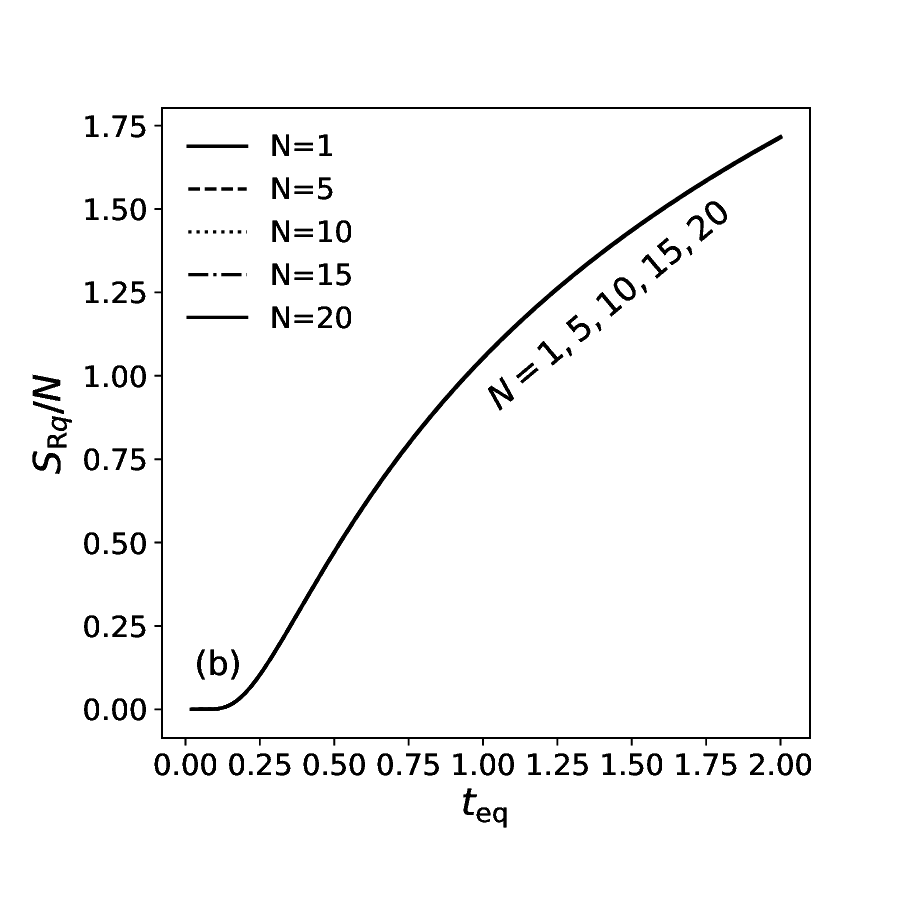}
    \label{fig:RenyiEntropy:Ndep:rescaled}
  }
  \caption{The R\'enyi entropies $\SR$ and the R\'enyi entropies divided by $N$ ($\SR/N$), as functions of $\teq$ at $q=0.98$ for $N=1, 5, 10, 15$, and $20$.}
\end{figure}

\begin{figure}
  \centering
  \subfigure[The R\'enyi entropies]
  {
    \includegraphics[width=0.45\textwidth]{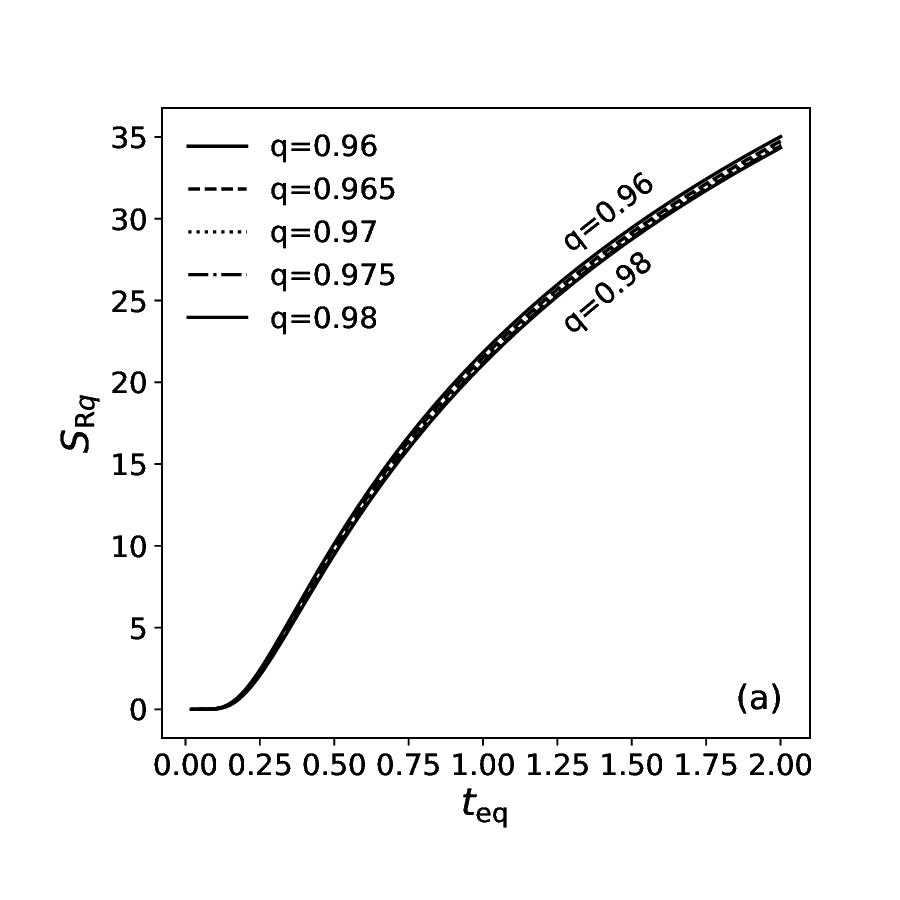}
    \label{fig:RenyiEntropy:qdep}
  }
  \hfill
  \subfigure[The R\'enyi entropies in the narrow range]
  {
    \includegraphics[width=0.45\textwidth]{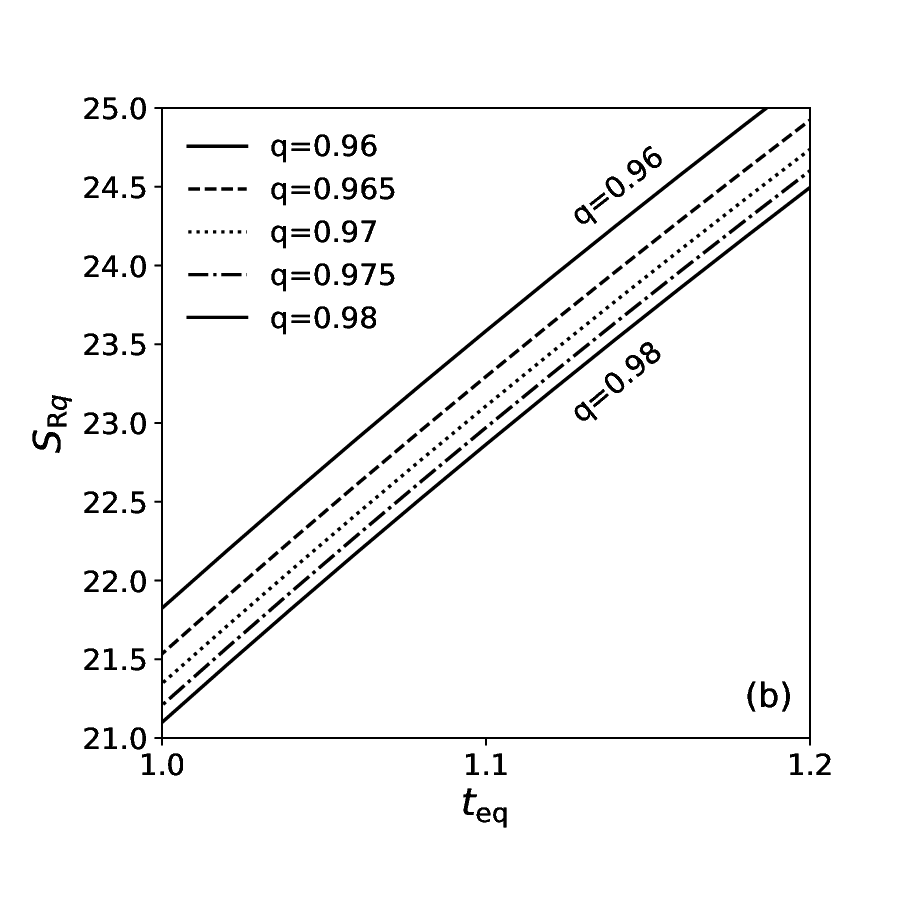}
    \label{fig:RenyiEntropy:qdep:narrow}
  }
  \caption{The R\'enyi entropies $\SR$ as functions of $\teq$ at $N=20$ for $q=0.96, 0.965, 0.97, 0.975$, and $0.98$.}
\end{figure}

Third, we calculate the R\'enyi entropy $\SR$ numerically from the Tsallis entropy $\ST$.
Figure~\ref{fig:RenyiEntropy:Ndep} shows the R\'enyi entropies $\SR$ at $q=0.98$ as functions of $\teq$ for $N=1, 5, 10, 15$, and $20$.
Figure~\ref{fig:RenyiEntropy:Ndep:rescaled} shows the R\'enyi entropies per oscillator $\SR/N$ at $q=0.98$
as functions of $\teq$ for $N=1, 5, 10, 15$, and $20$.
The R\'enyi entropy increases with $N$.
The $N$ dependence of the R\'enyi entropy per oscillator is exceedingly weak.
The R\'enyi entropy for $N$ oscillators is approximately $N$ times the R\'enyi entropy for a single oscillator. 
Figure~\ref{fig:RenyiEntropy:qdep}~\ref{fig:RenyiEntropy:qdep:narrow} shows the R\'enyi entropies 
$\SR$ at $N=20$ as functions of $\teq$ for $q=0.96, 0.965, 0.97, 0.975$, and $0.98$.
The range of Figure~\ref{fig:RenyiEntropy:qdep:narrow} is narrow. 
The $q$ dependence of the R\'enyi entropy is exceedingly weak. 

\begin{figure}
  \centering
  \subfigure[The heat capacities]
  {
    \includegraphics[width=0.45\textwidth]{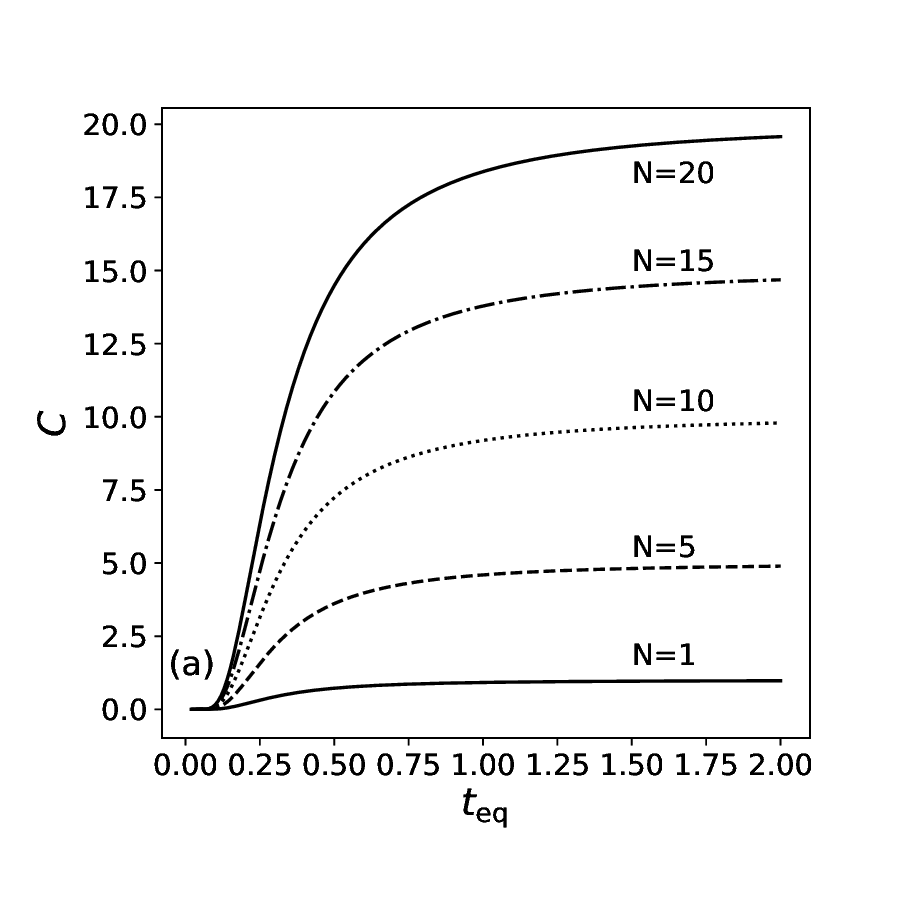}
    \label{fig:Ndep:HeatCapacity}
  }
  \hfill
  \subfigure[The heat capacities divided by $N$]
  {
    \includegraphics[width=0.45\textwidth]{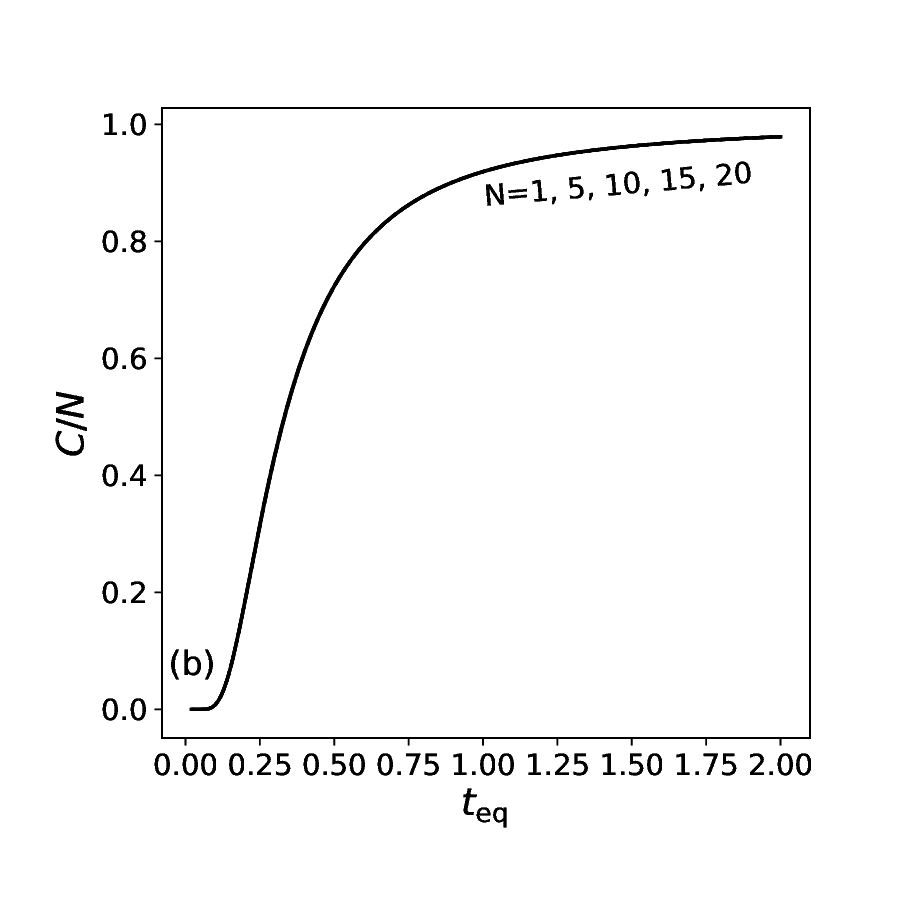}
    \label{fig:Ndep:HeatCapacity_div_N}
  }
  \caption{The heat capacities $C$ and the heat capacities divided by $N$ ($C/N$), as functions of $\teq$ at $q=0.98$ for $N=1, 5, 10, 15$, and $20$.}
\end{figure}

\begin{figure}
  \centering
  \subfigure[The heat capacities]
  {
    \includegraphics[width=0.45\textwidth]{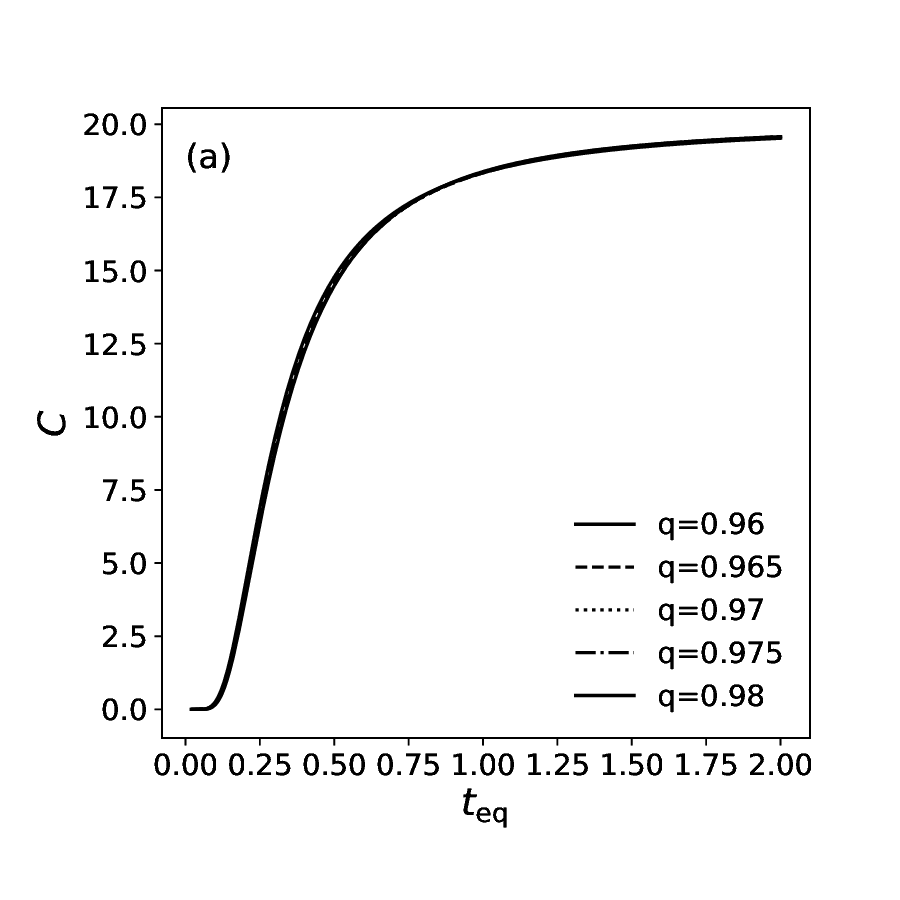}
    \label{fig:qdep:HeatCapacity}
  }
  \hfill
  \subfigure[The heat capacities in the narrow range]
  {
    \includegraphics[width=0.45\textwidth]{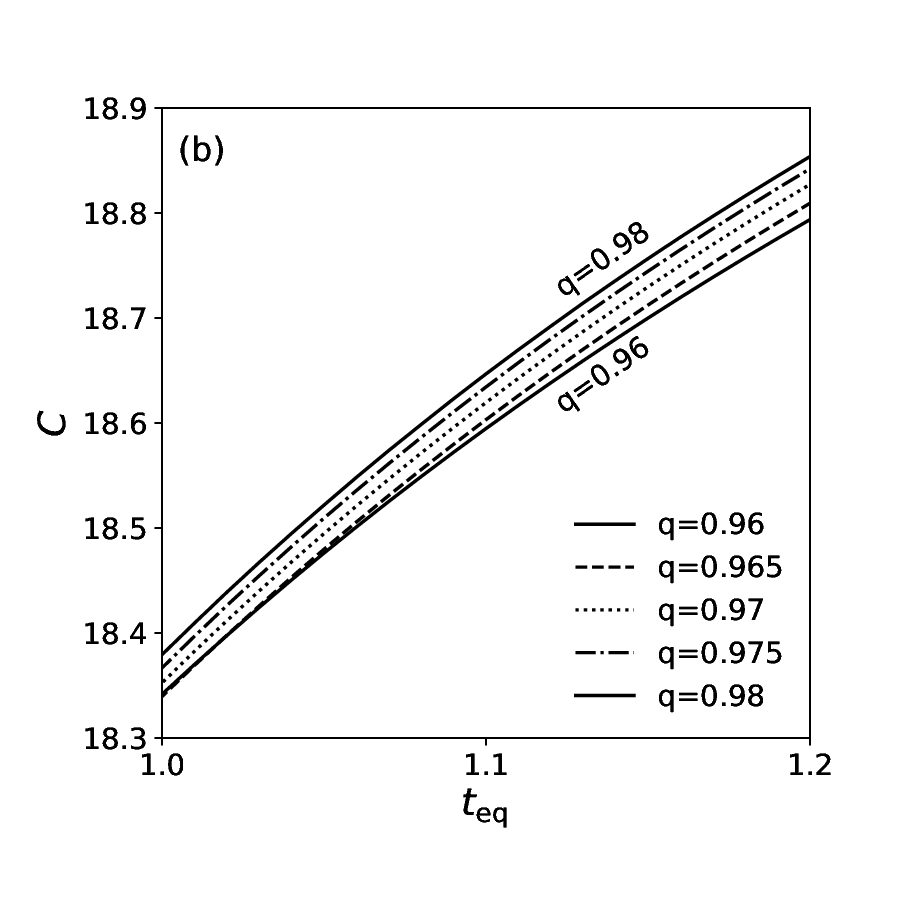}
    \label{fig:qdep:HeatCapacity:narrow}    
  }
  \caption{The heat capacities $C$ as functions of $\teq$ at $N=20$ for $q=0.96, 0.965, 0.97, 0.975$, and $0.98$.}
\end{figure}

Finally, we calculate the heat capacity $C$ numerically.
Figure~\ref{fig:Ndep:HeatCapacity} shows the heat capacities as functions of $\teq$ at $q=0.98$ for $N=1, 5, 10, 15$, and $20$.
The heat capacity increases with $N$.
Figure~\ref{fig:Ndep:HeatCapacity_div_N} shows the heat capacities per oscillator as functions of $\teq$ at $q=0.98$ for $N=1, 5, 10, 15$, and $20$.
The $N$ dependence of $C/N$ is exceedingly weak. 
The heat capacity for $N$ oscillators is approximately $N$ times the heat capacity for a single oscillator.
Figure~\ref{fig:qdep:HeatCapacity} \ref{fig:qdep:HeatCapacity:narrow} shows the heat capacities as functions of $\teq$ at $N=20$ for $q=0.96, 0.965, 0.97, 0.975$, and $0.98$.
The range of Fig.~\ref{fig:qdep:HeatCapacity:narrow} is narrow.
These figures indicate that the $q$ dependence of the heat capacity is exceedingly weak.
These dependences of the heat capacity reflect the dependences of the energy. 

\subsection{Probability distributions as functions of the energy}
We attempt to calculate numerically the probability distributions as functions of the energy for various values of the parameters. 
The probability distribution given by Eq.~\eqref{eqn:probdist_of_E} is shown using the solution of Eq.~\eqref{eqn:x:U}.
We use the parameter $M$ which equals $\ER/\aeq$ in the following calculations: 
the relation between $E$ and $M$ is $E=\aeq M + \sum_{j=1}^N b_j$. 
The data points are given for the values of $M$ which are non-negative integers. 
We provide figures to show distributions explicitly by removing markers for data points.

\begin{figure}
  \begin{center}
    \includegraphics[width=0.5\textwidth]{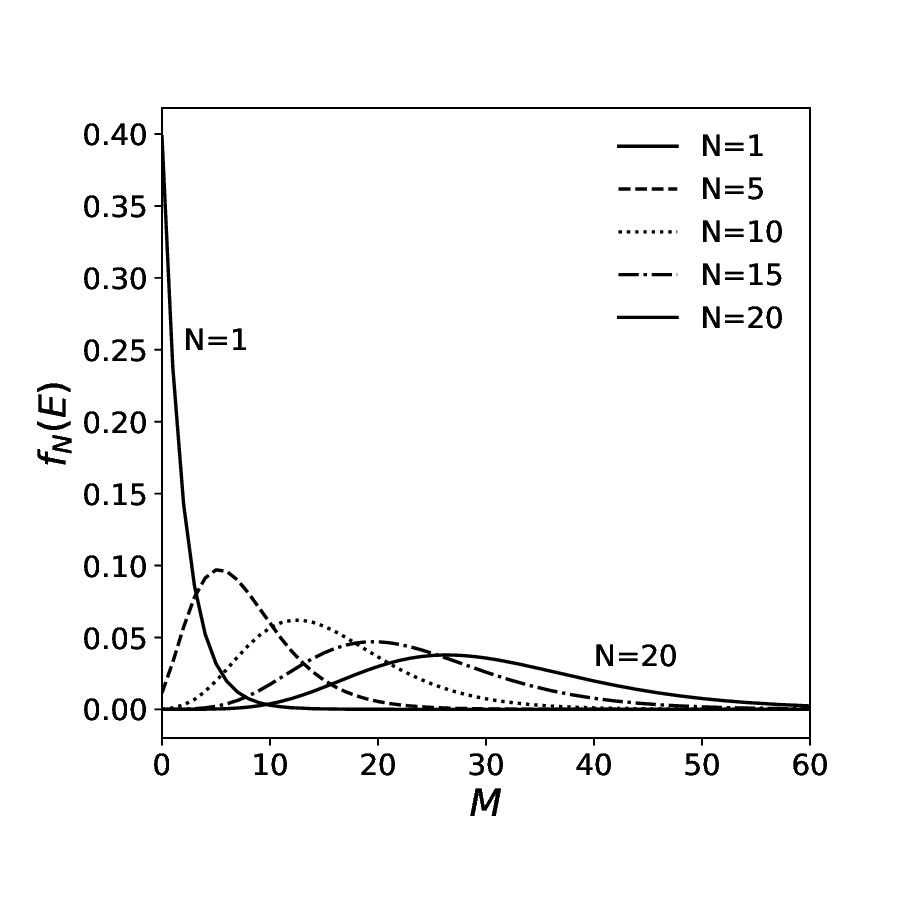}    
  \end{center}
  \caption{
    The probability distribution $f_N(E)$ at $q=0.8$ and $\teq=2.0$ for $N=1, 5, 10, 15$, and $20$.
    The data points are given for the values of $M$ which are non-negative integers.
    The markers for data points are removed from the figure.
  }
  \label{fig:dist:Ndep_without_points}
\end{figure}

First, we show the $N$ dependence of the probability distribution $f_N(E)$.
Figure~\ref{fig:dist:Ndep_without_points} shows $f_N(E)$ at $q=0.8$ and $\teq=2.0$ for $N=1, 5, 10, 15$, and $20$.
The distributions have tails, as expected. 
The $N$ dependence of the distribution is clearly seen, and
the value of $M$ at the peak of the distribution increases as the parameter $N$ increases.

\begin{figure}
  \begin{center}
    \includegraphics[width=0.5\textwidth]{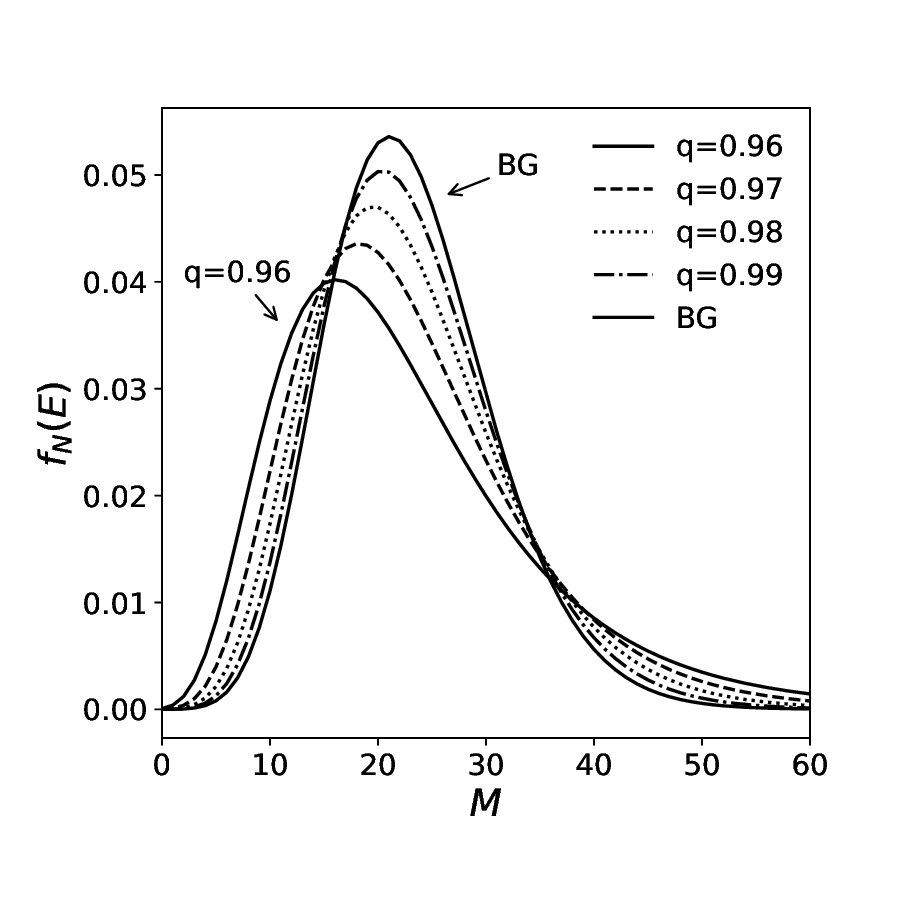}
  \end{center}
  \caption{
    The probability distribution $f_N(E)$ at $N=15$ and $\teq=2.0$ for $q=0.96, 0.97, 0.98$, and $0.99$.
    The label 'BG' indicates the distribution in the limit $q \rightarrow 1$,
    which corresponds to the distribution in the Boltzmann--Gibbs statistics. 
    The data points are given for the values of $M$ which are non-negative integers.
    The markers for data points are removed from the figure.
  }
  \label{fig:dist:qdep_without_points}
\end{figure}

Second, we show the $q$ dependence of the probability distribution $f_N(E)$. 
Figure~\ref{fig:dist:qdep_without_points} shows $f_N(E)$ at $N=15$ and $\teq=2.0$ for $q=0.96, 0.97, 0.98$, and $0.99$.
For comparison, we plot the probability distribution given by Eq.~\eqref{eqn:prob_ene_BGlim},
which corresponds to the distribution in the Boltzmann--Gibbs statistics.
The tail of the distribution becomes longer as $q$ decreases,
and the probability at low energy (small $M$) increases as $q$ decreases.
Therefore, as shown in Fig.~\ref{fig:dist:qdep_without_points}, 
the value of the probability at the peak of the distribution becomes lower as $q$ decreases.
The $q$ dependence of the distribution is evident in the figure. 

\begin{figure}
  \begin{center}
    \includegraphics[width=0.5\textwidth]{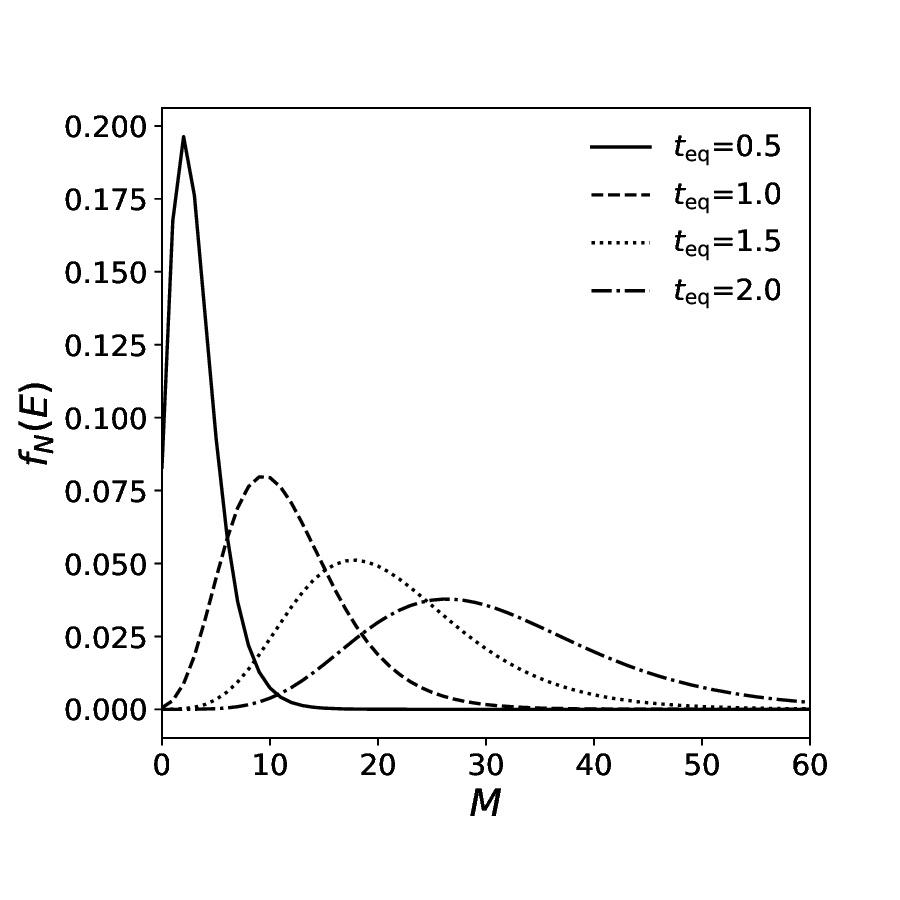}
  \end{center}
  \caption{
    The probability distribution $f_N(E)$ at $q=0.98$ and $N=20$ for $\teq=0.5, 1.0, 1.5$, and $2.0$.
    The data points are given for the values of $M$ which are non-negative integers.
    The markers for data points are removed from the figure.
  }
  \label{fig:dist:teqdep_without_points}
\end{figure}

Finally, we show the $\teq$ dependence of the probability distribution $f_N(E)$.
Figure~\ref{fig:dist:teqdep_without_points} shows $f_N(E)$ at $q=0.98$ and $N=20$ for $\teq=0.5, 1.0, 1.5$ and $2.0$.
The value of $M$ at the peak becomes larger and the width of the distribution becomes broader, as $\teq$ increases.
The value of the probability at the peak of the distribution becomes lower as $\teq$ increases. 
As shown in the figure, the distributions have tails.

\section{Discussion and Conclusions} 

We studied the physical quantities within the framework of the Tsallis statistics
employing the conventional expectation value (the linear average),
adopting the equilibrium temperature which is often called the physical temperature.
We treated power-law-like distributions: the entropic parameter $q$ is less than one.
We obtained the probability represented by the equilibrium temperature in the canonical ensemble under the maximum entropy principle.
We studied the thermodynamic quantities, and studied the probability distribution as a function of the energy. 
We obtained the expressions of the energy, the Tsallis entropy, and the heat capacity for $N$ harmonic oscillators,
where $N$ is the number of the oscillators.
We obtained the expressions of these quantities and calculated numerically these quantities,  
when the differences between adjacent energy levels are the same.
The R\'enyi entropy was also calculated from the Tsallis entropy. 
We calculated numerically the probability distributions for $N$ harmonic oscillators under such conditions. 

The derived probability is represented with the equilibrium temperature $\Teq$ and does not contain the Tsallis entropy,
whereas the probability derived in the previous studies is represented
with the temperature $T$ and contains the entropy, where $T$ is the inverse of the Lagrange multiplier.
The power exponent of the probability is $1/(q-1)$. 
For a power-law-like distribution, the parameter $q$ is less than one  in the present statistics,
whereas $q$ is larger than one
in both the Tsallis statistics employing the unnormalized $q$-expectation value
and the Tsallis statistics employing the normalized $q$-expectation value.
The region of the parameter $q$ for the power-law-like distribution depends on the definition of the expectation value. 

The $N$ dependences of the energy per oscillator, the R\'enyi entropy per oscillator, and the heat capacity per oscillator are exceedingly weak.
The $q$ dependences of the energy, the R\'enyi entropy, and the heat capacity are also exceedingly weak.
In contrast, the Tsallis entropy per oscillator depends on $N$ and the Tsallis entropy depends on $q$.
These behaviors are to be expected, 
when the quantity in the canonical ensemble is approximately equal to that in the microcanonical ensemble.
In the microcanonical ensemble, 
the Boltzmann--Gibbs entropy is given by $\ln W$, while the Tsallis entropy is given by $\ln_q W$,
where $W$ is the number of states and $\ln_q$ represents the $q$-logarithmic function which is defined by $\ln_q x = (x^{1-q} - 1)/(1-q)$. 
The R\'enyi entropy is also given by $\ln W$, which is shown by using the expression of the Tsallis entropy.
The relation between the R\'enyi entropy and the Tsallis entropy indicates
that the equilibrium temperature $\Teq$ corresponds to the temperature of the R\'enyi entropy $\SR$: $1/\Teq = \partial \SR/\partial U$.
When the equilibrium temperature is adopted,
it is expected that the $q$ dependences of the energy, the R\'enyi entropy, and the heat capacity in the canonical ensemble are weak,
provided that the thermodynamic quantity in the canonical ensemble is approximately equal to that in the microcanonical ensemble.

The probability distribution represented by the equilibrium temperature as a function of the energy has a tail,
and the distribution depends explicitly on $N$ and $q$.
As either $N$ or $\teq$ increases, or as both increases, 
the width of the distribution becomes wider,
the height of the peak of the distribution decreases,
and the position of the peak of the distribution moves to higher energy.
As $q$ decreases, the tail of the distribution becomes longer, the probability at low energy increases, 
and the position of the peak of the distribution moves to lower energy. 
The probability distribution depends on the entropic parameter $q$
when the distribution is represented by the equilibrium temperature.
The probability distribution is applicable to phenomena exhibiting power-law-like distributions.

We obtained the probability distribution represented by the equilibrium temperature
in the canonical ensemble within the framework of the Tsallis statistics employing the conventional expectation value.
The obtained results for the system of harmonic oscillators imply that the formalism is applicable to
a wide range of physical systems exhibiting power-law-like behavior.
We hope that the results are helpful for studies within the framework of unconventional statistical mechanics.  

\bigskip


\noindent
\textbf{CRediT authorship contribution statement}\\
\textbf{M.~Ishihara}
Conceptualization, Methodology, Formal analysis, Investigation, Software,
Validation, Visualization, Writing - original draft, Writing - review \& editing.

\medskip
\noindent
\textbf{Funding}\\
This research received no specific grant from any funding agency in the public, commercial, or not-for-profit sectors.

\medskip
\noindent
\textbf{Declaration of competing interest}\\
The author declares no conflict of interest.

\medskip
\noindent
\textbf{Data Availability}\\
This study is theoretical, and all numerical results can be reproduced from the equations provided in this paper.


\appendix
\section{Derivations of the probability and the density operator}
\label{appendix:derivations}
\subsection{Derivation of the probability within the framework of the Tsallis statistics employing the conventional expectation value}
\label{Tsallis-1-Parvan:probability}
The probability $p_i$ plays an essential role in classical statistical physics.
We apply the maximum entropy principle, and 
derive the probability $p_i$ within the framework of the Tsallis statistics employing the conventional expectation value.

The Tsallis entropy $\ST$ is given by 
\begin{align}
  \ST = \frac{\left( \sum_i (p_i)^q \right) - 1 }{1-q} . 
\end{align}
The functional with both the normalization condition and the energy constraint is given by 
\begin{align}
  I_{\mathrm{cl}} = \ST - \alpha \left[ \left( \sum_i p_i\right)  - 1 \right] - \beta \left[ \left(\sum_i p_i E_i \right) - U \right],
\end{align}
where $\alpha$ and $\beta$ are the Lagrange multipliers, $E_i$ is the value of the energy in state $i$, and $U$ is the energy.
The requirement $\delta I_{\mathrm{cl}} =0$ gives the equation:
\begin{align}
  \left(\frac{q}{1-q}\right) (p_i)^{q-1} - \alpha - \beta E_i = 0.
\end{align}
Therefore, we have 
\begin{align}
  p_{i} =  \left[\left( \frac{1-q}{q} \right)(\alpha + \beta E_i)\right]^{\frac{1}{q-1}}. 
  \label{tmp:prob}
\end{align}

From the definition of $\ST$, we have 
\begin{align}
\sum_i (p_i)^q  = 1 + (1-q) \ST . 
\end{align}
We calculate $\sum_i (p_i)^q$ by using Eq.~\eqref{tmp:prob} and obtain 
\begin{align}
\alpha = \left( \frac{q}{1-q} \right) [ 1 + (1-q) \ST ] - \beta U. 
\end{align}
We finally obtain
\begin{align}
  p_{i} =  \left[1 + (1-q) \ST + \left( \frac{1-q}{q} \right) \beta (E_i - U) \right]^{\frac{1}{q-1}}. 
\end{align}

\subsection{Derivation of the density operator within the framework of the Tsallis statistics employing the conventional expectation value}
\label{Tsallis-1-Parvan:density-operator}
The density operator $\hrho$ plays an essential role in quantum statistical physics.
We apply the maximum entropy principle, and 
derive the density operator $\hrho$ in the Tsallis statistics employing the conventional expectation value.

The Tsallis entropy $\ST$ is defined by 
\begin{align}
\ST = \frac{\Tr(\hrho^q) - 1}{1-q},  
\label{def:qm:rho}
\end{align}
where $\hrho$ is the density operator. 
The functional $I_{\mathrm{qm}}$ with both the normalization condition and the energy constraint is given by 
\begin{align}
I_{\mathrm{qm}} = \ST - \alpha (\Tr(\hrho) - 1) - \beta (\Tr(\hrho \hH) - U ) , 
\end{align}
where $\alpha$ and $\beta$ are the Lagrange multipliers, $\hH$ is the Hamiltonian, and $U$ is the energy.

We attempt to obtain the density operator $\hrho$ under the maximum entropy principle.
We use the method called quantum analysis
\cite{Suzuki-commun-math, Suzuki-Review-MathPhys, Suzuki:JMathPhys, Suzuki-progress, Suzuki-IJMPC10, Suzuki:Book:QuantAnalysis, Ishihara:2020:PhysicaA543, Ishihara:2020:PhysicaA} 
to obtain the density operator \cite{Ishihara:2020:PhysicaA543, Ishihara:2020:PhysicaA}. 
With this method, the following relation is obtained:
\begin{align}
  (\hrho + \varepsilon (\delta \hrho))^q
  = \hrho^q  + \varepsilon \int_0^1 \ dt\  q (\hrho - t \hdelta{\hrho})^{q-1} (\delta \hrho) + O(\varepsilon^2) ,
\label{eqn:QuantAnaly:expansion}
\end{align}
where $\hdelta{\hat{A}}$ for $\hat{A}$ operates $\hat{B}$ as follows: 
\begin{align}
\hdelta{\hat{A}} \hat{B} = \hat{A} \hat{B} - \hat{B} \hat{A}.
\end{align}
We remind that $\hrho$ and $\delta \hrho$ do not commute in general.
Using Eq.~\eqref{eqn:QuantAnaly:expansion} and the permutation property of the trace, we have
\begin{align}
\delta \ST = \varepsilon \frac{q}{1-q} \Tr(\hrho^{q-1} (\delta \hrho)) .
\end{align}

The maximum entropy principle $\delta I_{\mathrm{qm}} =0$ gives
\begin{align}
\left(\frac{q}{1-q}\right) \hrho^{q-1} - \alpha - \beta \hH = 0 .
\end{align}
Therefore, we obtain
\begin{align}
\hrho = \left(\frac{1-q}{q}\right)^{1/(q-1)}  (\alpha + \beta \hH)^{1/(q-1)} .
\end{align}
It is possible to eliminate $\alpha$ by calculating $\Tr(\hrho^q)$:
\begin{align}
1 + (1-q) \ST = \Tr(\hrho^q) = \left(\frac{1-q}{q}\right) (\alpha + \beta U) .
\end{align}
We finally obtain
\begin{align}
\hrho = \left(1 + (1-q) \ST + \left(\frac{1-q}{q}\right) \beta (\hH-U) \right)^{1/(q-1)} .
\end{align}


\end{document}